\documentclass[aps,prl,twocolumn,showpacs]{revtex4}
\usepackage[dvips]{graphicx}
\usepackage{amsmath}
\usepackage{amsfonts}
\usepackage{amssymb}
\usepackage{dsfont} 
\usepackage[latin1]{inputenc}
\usepackage{ulem}

\begin{document}

\title{Evaporative cooling of a small number of atoms\\ in a single-beam microscopic dipole trap}

\author{R.~Bourgain, J.~Pellegrino, A.~Fuhrmanek, Y.R.P.~Sortais and A.~Browaeys}

\affiliation{Laboratoire Charles Fabry, Institut d'Optique, CNRS, Univ Paris Sud,
2 Avenue Augustin Fresnel,
91127 PALAISEAU Cedex, France}

\date{\today}

\begin{abstract}
We demonstrate experimentally the evaporative cooling of a few hundred rubidium $87$ atoms in a single-beam microscopic dipole trap. Starting from $800$ atoms at a temperature of $125~\mu$K, we produce an unpolarized sample of $40$ atoms at $110$~nK, within $3$~s. The phase-space density at the end of the evaporation reaches unity, close to quantum degeneracy. The gain in phase-space density after evaporation is $10^3$. We find that the scaling laws used for much larger
numbers of atoms are still valid despite the small number of atoms involved in the evaporative cooling process. We also compare our results to a simple kinetic model describing the evaporation process and find good agreement with the data.
\end{abstract}

\maketitle

\section{Introduction}

Mesoscopic atomic ensembles containing a few hundred ultra-cold atoms constitute an interesting system for fundamental studies as well as applications in quantum optics and  atomic physics. For example, they have been proposed as a tool  to observe non-linear effects at the single photon level, e.g. using Rydberg atomic states~\cite{Saffman2002}, and more generally for applications in quantum information~\cite{Lukin01,Saffman10}. Recent proof-of-principle experiments along this line have been performed using samples containing $\sim 100$ laser-cooled atoms held in a microscopic dipole trap~\cite{Maxwell2013} or in one-dimensional
optical lattices~\cite{Dudin2012}. As another example, a quasi-deterministic single-atom source has been recently demonstrated using light-assisted collisions in a small cold atomic ensemble~\cite{GrunzweigNatPhys2010}. These demonstrations have been achieved with atomic samples laser cooled around the Doppler temperature, but it is often desirable to start from much colder atoms to decrease the sources of decoherence during quantum manipulations. Moreover, when further cooled down, these systems could allow the study of quantum degenerate gases in a regime where the number of atoms is small. An important experiment in this direction has been performed recently with fermions held in a tight dipole trap~\cite{Serwane2011}.

Several strategies can be used to prepare small ultra-cold atomic ensembles with temperatures in the micro-Kelvin range. One of them relies on the production of a macroscopic quantum degenerate or nearly-degenerate gas as a preliminary step; the ultra-cold sample is then used as a source to load a dipole trap in a dimple configuration, as demonstrated in Refs.~\cite{StamperKurn1998,Jacob2011} for large atomic samples and in Ref.~\cite{Serwane2011} for a few tens of fermionic atoms. An alternate route by-passes the first step mentioned above and starts direcly with a small sample of laser-cooled atoms confined in a tight trap of typically several micrometer size, such as magnetic traps on an atom chip~\cite{Whitlock2009} or arrays of optical dipole traps~\cite{Sebby2005} 
Evaporative cooling then allows to reduce the temperature (see e.g.~\cite{Luiten1996} and~\cite{Ketterle1996} and references therein) and has led so far to condensed samples containing several thousand atoms.

In the present work, we follow the second route and extend evaporative cooling to low atom numbers, ranging from a few hundreds to a few tens. Here, we use unpolarized rubidium atoms held in a single-beam microscopic dipole trap. Despite its simplicity, we show that this configuration is compatible with large elastic collision rates and leads to efficient evaporation. To do so, we lower the trap depth, as originally demonstrated by Adams {\it et al.}~\cite{Adams1995} and now routinely implemented in many laboratories to achieve quantum degenerate gases (see e.g.~\cite{Barrett2001,Granade2002,Cennini2003a,Weber2003,Kinoshita2005,Dumke2006,Gericke2007,Lauber2011}), by merely decreasing the power of the trap laser. We observe that the phase-space density increases by three orders of magnitude during the evaporation and eventually reaches unity, with clouds containing as few as $40$ atoms at $110$~nK. We also find that despite the small number of atoms involved, the scaling laws that govern the evolution of the thermodynamical quantities, derived by O'Hara {\it et al.}~\cite{OHara2001}, are still valid. Finally we show that a simple kinetic model of the evaporative cooling process inspired by the works of Refs.~\cite{Luiten1996,OHara2001,Comparat2006} reproduces our data well.

The paper is organized as follows. In Section~I, we describe our set-up and the experimental procedures. In particular, we detail our
strategy to load efficiently the microscopic dipole trap with up to $800$ atoms at a temperature around $125~\mu$K. Section~II shows our results on the evaporative cooling and compares the evolution of the temperature and the phase-space density to the scaling laws. In Section~III, we compare our data to the model of Refs.~\cite{Luiten1996,OHara2001,Comparat2006}. We conclude on possible improvements of the evaporative cooling in single-beam based experiments.

\section{I. Experimental set-up and procedures}\label{sec:setup}

\begin{figure}
\includegraphics[width=8cm]{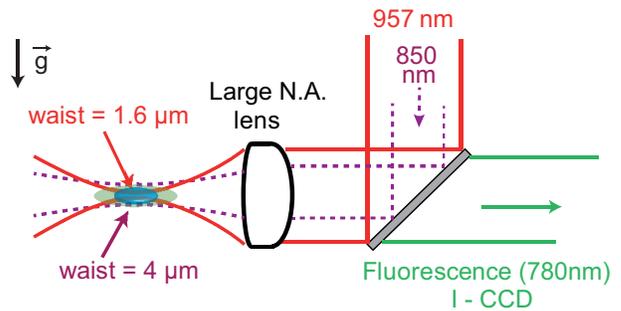}
\caption{(Color on-line) Trapping and detection scheme showing the two super-imposed single-beam dipole traps (see text) with comparable depths ($\sim 1$~mK) but different sizes (waists $4~\mu$m and $1.6~\mu$m). The large-numerical-aperture lens collects light-induced fluorescence. Imaging with an intensified camera (I-CCD) enables atom counting and temperature measurements, from which we extract the phase-space density. $\vec{g}$ indicates the direction of gravity.}
\label{fig:setup}
\end{figure}

Our set-up is sketched in Fig.~\ref{fig:setup}. We trap laser-cooled rubidium $87$ atoms in a microscopic dipole trap produced by focusing a laser beam at $957$~nm with a large-numerical-aperture aspheric lens, as described in Ref.~\cite{Sortais2007}. The $1/e^2$ radius of the gaussian spot is $w=1.6~\mu$m. The size of the trap along the longitudinal direction is characterized by the Rayleigh length, $z_{\rm R} = 8.4$~$\mu$m. We use $20$~mW of laser power to achieve a trap depth $U/k_{\rm B} = 1$~mK ($k_{\rm B}$ is the Boltzman constant). For this depth, the trapping frequencies of the atoms in the transverse and longitudinal directions are $\omega_\perp/2\pi=64$~kHz and $\omega_\parallel/2\pi=9$~kHz respectively. For a temperature of $125~\mu$K, this corresponds to a thermal cloud with root-mean-square sizes $270$~nm and $2~\mu$m respectively.

For all the measurements reported here, we measure the number of atoms $N$ after the cloud has been released in free space. For that purpose, we send a $10~\mu$s pulse of circularly-polarized probe light resonant with the $|5S_{1/2},F=2,M=2\rangle$ to $|5P_{3/2},F'=3,M'=3\rangle$ transition, combined to repumping light
tuned to the $(5S_{1/2},F=1)$ to $(5P_{3/2},F=2)$ transition. The intensity of the probe beam is $I/I_{\rm sat}=1$ ($I_{\rm sat}=1.6$~mW/${\rm cm}^2$). The large-numerical-aperture aspheric lens collects the light-induced fluorescence, which we detect with an image intensifier followed by a low-noise charge-coupled device camera (I-CCD). The duration of the time-of-flight is chosen long enough for the density to drop below $10^{11}$ at/${\rm cm}^3$ to avoid light-assisted losses during probing~\cite{Fuhrmanek2012}. In this way, the detected fluorescence is proportional to the number of atoms, which we extract by calibrating our detection system with a single atom~\cite{Fuhrmanek2010}. The small number of atoms $N$ requires that we integrate the signal over several realizations of the experiment, typically $10$ to $1000$ for $N$ ranging from $800$ to $40$, allowing us to determine $N$ with a typical statistical uncertainty of $10\%$. We measure the temperature $T$ of the cloud with the time-of-flight technique with a statistical uncertainty of $10\%$.

Efficient evaporative cooling requires a large elastic collision rate and therefore that we initially confine a large number of atoms in the microscopic dipole trap. To do so, we proceed in four steps : first, we load a magneto-optical trap (MOT) in $1$~s using a Zeeman slower. In a second step, we load from this MOT a dipole trap with a $1/e^2$ radius of $4~\mu$m and a depth of $1.4$~mK (see fig.~\ref{fig:setup}) produced by focusing a laser beam at $850$~nm with the above-mentioned aspheric lens (the full aperture of the lens is not used here). For this purpose, we use a compressed MOT sequence during which we reduce the intensity of the MOT lasers whilst red-detuning them from the fluorescence resonance. This results in a cloud of up to $3500$ atoms at a temperature of $200~\mu$K and in a mixture of Zeeman sub-levels $M=0,\pm 1$ of the $F=1$ hyperfine ground state. In a third step, we switch off the MOT lasers and turn on the microscopic dipole trap ($w=1.6~\mu$m, depth of $1$~mK), which acts as a dimple. Finally, after 200~ms, we switch off the $4~\mu$m-size trap, and following $60$~ms of plain evaporation, $\sim20\%$ of the atoms initially in the $4~\mu$m-size trap are left in the microscopic dipole trap.

At the end of the loading procedure, $N\approx800$ atoms are trapped in the $U/k_{\rm B}=1$~mK deep microscopic trap, at a temperature $T\approx125$~$\mu$K. The atoms are prepared in the $F=1$ hyperfine ground state level, in a mixture of  Zeeman sub-states. Assuming a deep harmonic trap, the thermal volume is  $V_{{\rm th}}=(2\pi k_{\rm B} T /(m\omega^2))^{3/2}$, corresponding to a spatial density at the center of the trap $n_0=N/V_{{\rm th}}\approx3\times10^{14}$~at.${\rm cm}^{-3}$ and an initial phase-space density $D=N (\hbar \omega/k_{\rm B} T)^3 \approx1.5\times 10^{-3}$. Here, $\omega=(\omega_\perp^2 \omega_\parallel)^{1/3}$ is the mean oscillation frequency of the trap. Note that the phase-space density calculated in this way is based on the total number of atoms, irrespective of their Zeeman sub-level. In the present status of the experiment we do not measure the population in each Zeeman states. We calculate the  elastic collision rate  $\gamma_{\rm el}=n_{0}\sigma \bar{v}\sqrt{2}\approx 3\times 10^4\ {\rm s}^{-1}$. Here, $\bar{v}=\sqrt{8 k_{\rm B} T/(\pi m)}$ is the thermal average velocity ($m$ is the atomic mass). The effective elastic cross-section is $\sigma =\epsilon~4\pi a^2$ with $\epsilon=2~{\rm or}~4/3$ for, respectively, atoms all in the same Zeeman sub-level of $F=1$, or atoms in an equal mixture of Zeeman sub-states. Here, $a=5$~nm is the scattering length.

\section{II. Experimental results and comparison to scaling laws}\label{sec:exp}

\begin{figure}
\includegraphics[width=8cm]{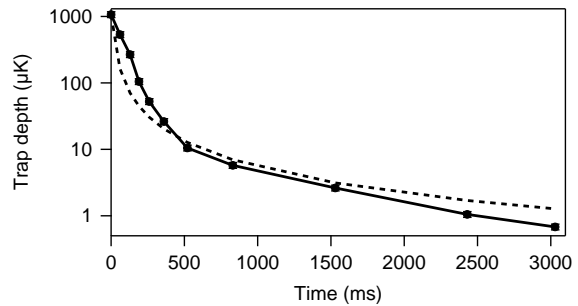}
\caption{(Color on-line) Evolution of the trap depth as a function of time during the evaporation. Solid line: experimental ramp with linear pieces after optimization. The trap depth is measured with a $10\%$ uncertainty. Dotted line: theoretical prediction by O'Hara \textit{et al.}~\cite{OHara2001}.}
\label{fig:rampe_evaporation}
\end{figure}

Once the microscopic dipole trap has been loaded, we apply forced evaporation by decreasing the  power of the dipole trap beam using an acousto-optical modulator. We calculate the depth $U(t)$, assuming a Gaussian laser beam and taking into account the deformation of the dipole potential by gravity. The evaporation ramp consists of $10$ pieces, each piece corresponding to a linear decrease of the trap depth.  At the end of each step, we maximize the phase-space density by adjusting the duration of the linear piece~\footnote{It could happen that a minor gain in phase-space resulted into an important loss of atoms. In this case we kept the ramp corresponding to the largest number of atoms.}. The experimental ramp obtained in this way is represented in fig.~\ref{fig:rampe_evaporation}.

For the sake of comparison we also plot the theoretical ramp predicted by O'Hara \textit{et al.}~\cite{OHara2001}, which is valid when the ratio $\eta = U/(k_{\rm B} T)$ remains constant throughout the evaporation. Also, this prediction holds in the absence of inelastic losses. In that case, the ramp is
$U(t)=U_{\rm i}{\left(1+t/\tau\right)^{\frac{2(3-\eta')}{\eta'}}}$, with $U_{\rm i}$ the initial trap depth, $\eta' = \eta + \kappa$ and $\kappa =(\eta-5)/(\eta-4)$ when $\eta \gg 1$. The time constant $\tau$ is related to $\eta$ and the initial elastic collision rate $\gamma_{\rm el,i}$ through $\tau=\left(\frac{2}{3}\eta'(\eta-4)e^{-\eta}\gamma_{\rm el,i}/(2\sqrt{2})\right)^{-1}$, on the order of $15$~ms, using our initial measured value $\eta\simeq 8.5$.

\begin{figure}
\includegraphics[width=8cm]{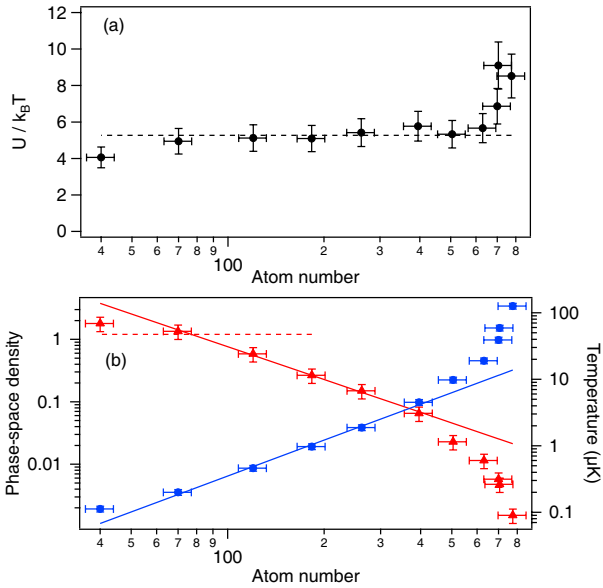}
\caption{(a) Evolution of the ratio $\eta=U/k_{\rm B} T$ versus the number of atoms $N$ during the evaporation (dots). Dashed line: average value of $\eta$ for $70\leq N\leq 400$. (b) Evolution of the measured temperature $T$ (dots, right axis) and the phase-space density $D=N (\hbar \omega/k_{\rm B} T)^3$ (triangles, left axis) versus the number of atoms. The solid lines are fits by power laws performed for the data corresponding to $70\leq N\leq 400$. Error bars correspond to $10\%$ uncertainties on the measured temperature, number of atoms, trap depth $U$, and oscillation frequency $\omega$. In (b), the dashed line at value $D=\zeta(3)\simeq1.202$ indicates the transition between a thermal and degenerate polarized sample.}
\label{fig:scalinglaws}
\end{figure}

Not surprisingly, the experimental ramp deviates from the above-mentioned theoretical prediction, especially in the early stage of the evaporation. This is due to the underlying assumptions of the prediction not being fulfilled in our case. Figure~\ref{fig:scalinglaws}a shows that $\eta$ decreases rapidly at the beginning of the evaporation, before it stabilizes around $5$. It is therefore not constant throughout the evaporation and not much larger than $1$. Moreover, in our case, the inelastic losses are not negligible at the beginning of the evaporation ramp (see Section~III). We also note that the initial elastic collision rate is on the order of the oscillation frequency $\omega_\parallel$ along the longitudinal axis of the trap. The cloud may therefore be partially in the hydrodynamics regime (see e.g.~\cite{Beijerinck2000}), an effect not taken into account here.

Figure~\ref{fig:scalinglaws}b shows the evolution of the temperature versus the atom number during the evaporation. Combining the temperature and the number of atoms measured at the end of each linear piece of the ramp to the knowledge of the trap depth $U$, we calculate the average oscillation frequency $\omega$ and the phase-space density $D=N (\hbar \omega/k_{\rm B} T)^3$ at the center of the trap assuming an infinitely deep harmonic trap. The evolution of $D$ is represented also in fig.~\ref{fig:scalinglaws}b: we observe a gain in phase-space density of $10^3$ during the evaporation. The data suggest that, assuming a polarized sample, we may have reached quantum degeneracy at the end of the ramp, as $D$ then exceeds the threshold value $\zeta(3)\simeq 1.202$ (with the usual definition of the Riemann function $\zeta(n)=\sum_{j=1}^\infty 1/j^n$) \cite{KetterlePRA1996}. However, we could not see any evidence of a double-structure on the column density after time-of-flight. This fact may be due to the small number of atoms involved~\cite{KetterlePRA1996}: at the threshold $k_{\rm B} T_{\rm c}\approx \hbar\omega~ (N/\zeta(3))^{1/3}\approx 3 \hbar\omega$. This implies that the thermal and the quantum degenerate components of the gas would have very similar sizes in a time-of-flight measurement (the ratio of the two sizes is $\sim \sqrt{k_{\rm B}T/(\hbar\omega)}$ in the absence of interaction between atoms). Furthermore, we do not have any access to the anisotropy of the cloud due to the geometry of the experiment, as we look along the long axis of the traps (see fig.~\ref{fig:setup}). More involved diagnostics will be necessary to study the quantum degenerate regime, such as stimulated Raman spectroscopy~\cite{Grynberg93,Kaufman2012,Thompson2013}, which we leave for future work.

As a first step to understand evaporation processes in our single-beam trapping configuration, we compare the data in fig.~\ref{fig:scalinglaws}b to the scaling laws derived by O'Hara \textit{et al.}~\cite{OHara2001}, which are valid for a constant ratio $\eta$ and in the absence of inelastic losses. Following~\cite{OHara2001},  the phase-space density should scale as $D\propto N^{-\xi}$, with $\xi=\eta'-4$ characterizing the efficiency of the evaporation~\cite{Ketterle1996}. In the same way, the temperature is expected to scale with the number of atoms as $T\propto N^{2(\eta'-3)/3}$. We have fitted the scaling laws to the part of the data for which $\eta$ is almost constant, i.e. $70\leq N\leq400$ (see fig.~\ref{fig:scalinglaws}a). We find that $d\ln T/d\ln N=1.79(6)$ and that the efficiency of the evaporation is $\xi=1.75(5)$. These two values lead to two independent and consistent evaluations of $\eta=5.4(1)$~\footnote{These two determinations of $\eta$ are independent, as the calculation of the phase-space density depends also on the value of the trap depth.}. In the above-mentioned range for $N$, the scaling laws are also consistent with the average ratio $\eta=5.3(3)$, which we directly deduce from $U$ and $T$ (see fig.~\ref{fig:scalinglaws}a). Thus, in as much as the ratio $\eta$ remains constant and the inelastic losses can be neglected (which is the
 case for $N\leq 400$, see Section~III), the scaling laws predicted by O'Hara \textit{et al.} are valid even for the small value $\eta\sim5$ and atom numbers as small as a few tens.

\section{III. Evaporation dynamics}\label{sec:dynamics}

In order to better understand the evaporation processes involved in our geometry, in particular the effect of the inelastic losses at the beginning of the evaporation process ($N>400$), we have simulated the evaporative cooling process using the kinetic model developed in Refs.~\cite{Luiten1996,OHara2001,Comparat2006}. This model assumes that the atomic cloud is in thermal quasi-equilibrium at any time during the evaporation, that it is described by a truncated Boltzman distribution, and that the atomic trajectories are ergodic. Based on these assumptions, one calculates thermodynamical quantities for the gas and relates them through rate equations. Here, we include in the model one-, two- and three-body inelastic losses, as well as heating due to spontaneous emission induced by the trapping laser, and we take into account the time-dependence of the parameter $\eta$.

We start with the average energy of the atomic cloud assumed to be trapped in a deep harmonic potential, $E = 3 N k_{\rm B} T$, which yields $\dot{E}/E=\dot{N}/N+\dot{T}/T$. Four mechanisms lead to a variation of the energy during the evaporation. The first one is evaporation due to elastic collisions between trapped atoms. Atoms leaving the trap have an average energy $(\eta + \kappa) k_{\rm B} T$, with $\kappa =(\eta-5)/(\eta-4)$~\footnote{This expression
of $\kappa$ holds only in the limit $\eta\gg 1$. To account for our not so large values of $\eta$ at the end of the evaporation ramp, we could use the general expression of $\kappa=1-P(5,\eta)/(\eta P(3,\eta)-4 P(4,\eta))$, where $P$ is the incomplete gamma function~\cite{Abramowitz1972}. We checked that this alters the overall results in $N$ and $T$ by less than a few percents.}. Assuming 3D-evaporation, this mechanism leads to a rate of variation in the energy $\dot{E}_{\rm ev} = (\eta + \kappa) k_{\rm B} T \dot{N}_{\rm ev}$, where the atom loss rate due to the evaporation is $\dot{N}_{\rm ev}/N = -2(\eta-4)~e^{-\eta}~\gamma_{\rm el}/(2\sqrt{2})$~\cite{Luiten1996,OHara2001}~\footnote{This expression of $\dot{N}_{\rm ev}/N$ holds for $\eta\gg 1$. We also checked that using the general expression $\dot{N}_{\rm ev}/N=-2(\eta P(3,\eta)-4 P(4,\eta))~e^{-\eta}~\gamma_{\rm el}/(2\sqrt{2})$ does not change significantly the result of the simulation.}. The second mechanism results from the decrease of the oscillation frequency due to the adiabatic reduction of the trap depth, leading to a rate $\dot{E}_{\rm ad}=E\, \dot{U}/(2U)$ for a harmonic trap. The third mechanism is the heating due to spontaneous emission induced by the dipole trap laser. We model the associated variation of energy as $\dot{E}_{\rm heat} = 2 E_{\rm r} R\, N$, where $E_{\rm r}$ is the recoil energy gained in  the scattering of a photon by an atom, which occurs at a rate $R(t)\propto U(t)$. Finally, the energy varies due to the loss of atoms by one-, two- or three-body inelastic collisions. Each $q-$body loss ($q=1,2$ or $3$) is governed by the equation on the local density $\dot{n}_{q}({\bf r},t)=-K_{q} n({\bf r},t)^q$. It gives a rate of variation in the energy
\begin{equation}
\dot{E}_{q} = \frac{3}{2}\, \dot{N}_{q}\,  k_{\rm B} T + \int U(\textrm{\bf r})\, \dot{n}_{q}(\textrm{\bf r},t)\,  d^3\textrm{\bf r}\ ,
\end{equation}
where the first and second terms are the contributions of the kinetic energy and of the trapping potential $U({\bf r})$, respectively~\cite{Luiten1996}. Here, $\dot N_{q}=\int \dot{n}_{q}(\textrm{\bf r},t) d^3\textrm{\bf r}$.
Plugging in the equation for the evolution of the local density yields:
\begin{equation}
\dot{E}_{q} = \left(\frac{3}{2}+\gamma_{q}\right)\, \dot{N}_{q}\,  k_{\rm B} T \ ,
\end{equation}
where
\begin{equation}
\gamma_{q} = \frac{1}{q} \, \frac{T}{V_{q}} \frac{dV_{q}}{dT}\ \textrm{and}\
V_{q}=\int e^{-q \frac{U(\textrm{\bf r})}{k_{\rm B} T}} d^3\textrm{\bf r}\ .
\end{equation}
For an infinitely deep harmonic trap, we find $\gamma_{q}=3/(2q)$, showing that two- and three-body losses carry away less than the average energy, leading to anti-evaporation. This is due to the fact that these losses occur mainly at the center of the trap, where the density is largest and the energy of the atoms is smaller than the average energy. The atom loss rate due to $q$-body inelastic collisions is $\dot{N}_{q}/N = - K_{q}n_{0}^{q-1}/(q\sqrt{q})$.

Using $\dot{N}=\dot{N}_{\rm ev}+\sum_{q}\dot{N}_{q}$ and $\dot{E}=\dot{E}_{\rm ev}+\dot{E}_{\rm ad}+\dot{E}_{\rm heat}+\sum_{q}\dot{E}_{q}$, we obtain the set of coupled equations governing the evaporation in the presence of heating and $q$-body losses:
\begin{eqnarray}\label{eq:diff_eqs}
 \frac{\dot{N}}{N}&=& -2(\eta-4) e^{-\eta}\frac{\gamma_{\rm el}}{2\sqrt{2}}+\sum_{q=1,2,3}\frac{\dot{N}_{q}}{N},\nonumber \\
 \frac{\dot{T}}{T}&=& -2\left(\frac{\eta+\kappa}{3}-1\right)(\eta-4) e^{-\eta}\frac{\gamma_{\rm el}}{2\sqrt{2}}\nonumber\\
 &+&\sum_{q=1,2,3}\frac{1}{3}\left(\frac{3}{2}+\gamma_{q}-3\right)\frac{\dot{N}_{q}}{N}\nonumber\\
 &+&\frac{2 E_{\rm r} }{3k_{\rm B} T}\, R+\frac{\dot{U}}{2U}\ ,
\end{eqnarray}
where $\eta$, $R$ and $\gamma_{\rm el}$ are functions of time. We solve these equations numerically using the experimental ramp $U(t)$ of fig.~\ref{fig:rampe_evaporation}. We find that the heating rate $R(t)$, which amounts initially to $R(0)=22$~s$^{-1}$, drops very rapidly and has very little influence on the evaporation. For the one-body loss rate, we plug in the simulation the value $K_{1}= 0.1$~s$^{-1}$ corresponding to the measured $10$~s vacuum-limited lifetime of a single atom in the microscopic trap~\cite{Sortais2007}. Also, we take for the three-body loss rate the value that we measured in a separate experiment, $K_{3}=4\times 10^{-29}$~${\rm cm}^6$.s$^{-1}$~\footnote{Our value $K_{3}=4\pm3\times 10^{-29}$~${\rm cm}^6$/s, measured on a cloud at a temperature of $100~\mu$K, is compatible with the values obtained in~\cite{Burt1997,Soding1999} at much lower temperatures. }. To reproduce our data more accurately we have to assume a two-body loss rate $K_{2}= 1.5\times10^{-14}$~cm$^3$.s$^{-1}$. The two-body losses may be due to a small steady population of atoms in state $F=2$ caused by near-resonant laser being improperly switched off. This population results into hyperfine changing collisions with the atoms in the $F=1$ level. Reference~\cite{Gensemer2000} reports a value for the hyperfine changing collision rate of $8\times10^{-12}$~cm$^3$.s$^{-1}$. We infer from this value a fraction of atoms in the $F=2$ states of $0.2\%$, too small to be measured directly with our set-up.

\begin{figure}
\includegraphics[width=8cm]{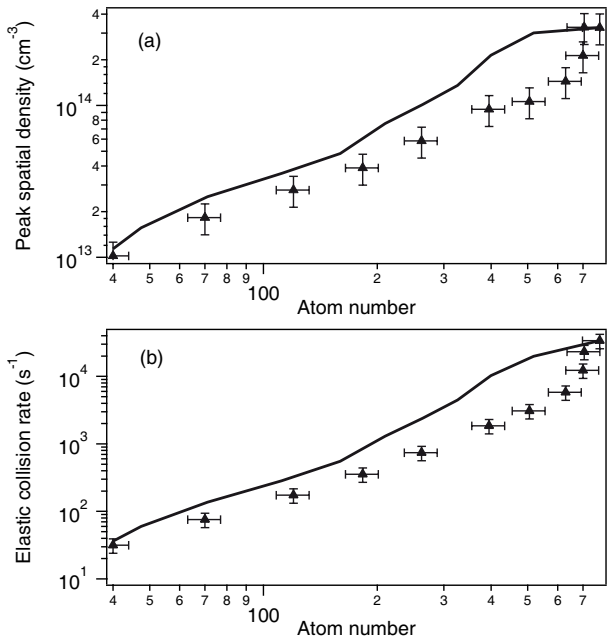}
\caption{Evolution of (a) the spatial density $n_{0}$ and (b) the elastic collision rate $\gamma_{\rm el}=n_{0}\sigma\bar v \sqrt{2}$ at the center of the trap, versus the number of atoms. Triangles: values extracted from the measurements of $T$, $N$ and the trap depth $U$. Error bars correspond to $10\%$ uncertainties on $T$, $N$ and the oscillation frequency $\omega$. Solid lines: kinetic model using $\sigma=\epsilon~4\pi a^2$ with $\epsilon=4/3$ (i.e. the atoms are equally distributed among the Zeeman sub-levels of $F=1$).}
\label{fig:evolution_thermodynamic}
\end{figure}

The contributions of the two- and three-body losses in eqs.~(\ref{eq:diff_eqs}) are significant essentially at the beginning of the ramp, where the loss rates amount to $K_{2}n_0/(2\sqrt{2})\simeq 1.7$~s$^{-1}$ and $K_{3}n_0^2/(3\sqrt{3})\simeq 0.8$~s$^{-1}$, respectively. As shown in fig.~\ref{fig:evolution_thermodynamic}a, the density drops below $10^{14}$~at.${\rm cm}^{-3}$
when the number of atoms becomes smaller than $400$. The inelastic losses thus become negligible. This justifies the experimental validity of the scaling laws detailed in Section~II. For $N\leq 400$, the main contributions to the evaporation process are the adiabatic reduction of the trap depth and the elastic collision rate, which has decreased by $\sim3$ orders of magnitude by the end of the ramp (see Fig.~\ref{fig:evolution_thermodynamic}b).

\begin{figure}
\includegraphics[width=8cm]{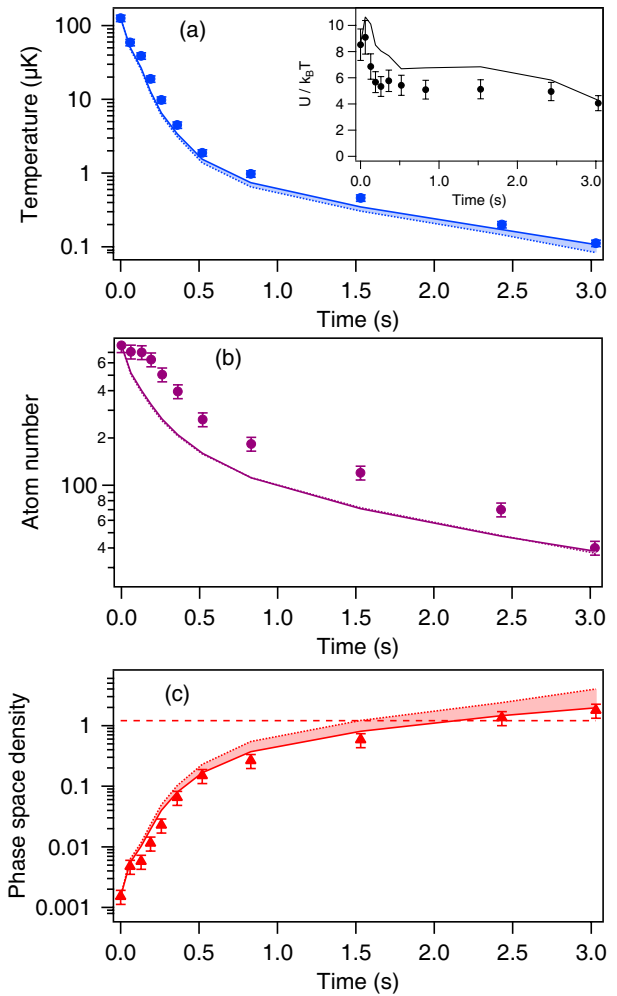}
\caption{Comparison between the data and the model of eqs.~\ref{eq:diff_eqs}. Evolution of (a) the temperature $T$, (b) the number of atoms $N$, and (c) the phase-space density $D=N (\hbar \omega/k_{\rm B} T)^3$, as a function of the evaporation time. The dots and triangles are the data. Error bars correspond to $10\%$ uncertainties on $T$, $N$ and the oscillation frequency $\omega$. The solid and dotted lines are respectively the predictions of the model for $\sigma=(4/3)\times 4\pi a^2$ (i.e. the atoms are equally distributed among the Zeeman sub-levels of $F=1$) and $\sigma=2\times 4\pi a^2$ (i.e. the atoms are all in the same Zeeman sub-level). Here, we have taken $\{K_1;K_2;K_3\}=\{0.1$~s$^{-1};1.5\times10^{-14}$~cm$^3$.s$^{-1};4\times 10^{-29}$~${\rm cm}^6$.s$^{-1}\}$. In (c), the dashed line corresponds to $D=\zeta(3)\simeq 1.202$. Inset in (a) : evolution of $\eta=U/k_BT$.}
\label{fig:comparaison_data_theo}
\end{figure}

Using the above-mentioned parameters for the heating and inelastic loss rates, eqs.~(\ref{eq:diff_eqs}) yield the evolution of the temperature, number of atoms and phase-space density as a function of the evaporation time. Figure~\ref{fig:comparaison_data_theo} compares the experimental data with the results of the model for values of the elastic cross-section $4/3\leq\sigma/(4\pi a^2)\leq 2$. The agreement between the simulation and the data is better than $30\%$ for the temperature and the parameter $\eta$ (see fig.~\ref{fig:comparaison_data_theo}a and inset). By setting $K_{2}=0$ and $K_{3}=0$ in the simulation, we have checked that the inelastic losses at the beginning of the ramp are largely responsible for the decrease of $\eta$ in the early stage of the evaporation.
The agreement between the model and the data is somewhat less satisfactory for the number of atoms (see fig.~\ref{fig:comparaison_data_theo}b). In particular, $N$ decreases experimentally more slowly than predicted at the beginning of the ramp. This could be due to the fact that the model does not account for the possible hydrodynamic behavior at the beginning of the ramp, as mentioned in Section~II. It could also come from the fact that evaporated atoms remain in the vicinity of the trap when we perform imaging of the cloud (and therefore contribute to the measured number of atoms), but do not collide with the remaining trapped atoms (and therefore do not contribute to the evaporation at a later time). Finally, fig.~\ref{fig:comparaison_data_theo}c shows good agreement between the simulation and the data for the phase-space density. The slowing down of the evaporation seen both in the experiment and the theory is a consequence of the decrease of the elastic collision rate when the trapping potential is lowered, even in the absence of inelastic losses (see e.g.~\cite{OHara2001}). In our case, inelastic losses further slow down the evaporation process at the beginning of the evaporation : in the absence of these losses (but yet with the same microscopic trapping potential and evaporation ramp), the phase space density would reach unity after $700$~ms only.

\section{Conclusion}

In conclusion, we have implemented evaporative cooling on a few hundred atoms held in a microscopic  dipole trap. After an evaporation period of $3$~s we reach a phase-space density of $\sim 2$ with $40$ unpolarized atoms at $\sim100$~nK, close to quantum degeneracy. The overall duty cycle of the experiment is around $4$ seconds. The atomic sample thus obtained is a good starting point for experiments involving mesoscopic-sized ultra-cold atomic samples. In the future, it will be interesting to enhance the efficiency of the evaporation by tilting the potential, e.g. by applying a strong magnetic field gradient~\cite{Hung2007,Serwane2011}, or an extra laser beam to achieve the runaway regime~\cite{Clement2009}. This may allow us to reach Bose-Einstein condensation with a small number of atoms, and to decrease the duration of the evaporation.

From a theoretical perspective, we have compared our data to a simple model of evaporative cooling, assuming the trapping potential to be harmonic. We have found that the scaling laws governing the evolution of the thermodynamics quantities are still relevant in our regime of small atom numbers, as long as the ratio $\eta$ remains constant and the inelastic losses are negligible. We have also found that a kinetic model taking into account time variations of $\eta$ as well as heating and loss mechanisms reproduces the dynamics of the evaporation in a fair way. This simple model is therefore a useful guide to optimize evaporation parameters.
In the future it will be interesting to refine the model presented here by taking into account the real shape of the Gaussian laser trap potential~\cite{Simon2010,Sortais2013}.

\begin{acknowledgments} We thank T.~Lahaye for careful reading of the manuscript. We acknowledge support from the E.U. through the ERC Starting Grant ARENA and the AQUTE Project, and from the Region Ile de France (Triangle de la Physique and IFRAF).
\end{acknowledgments}

\end{document}